\documentclass[12pt,preprint]{aastex}

\shorttitle{Rotochemical Heating of Neutron Stars}
\shortauthors{Reisenegger et al.}

\begin{document}
\title{Rotochemical Heating of Neutron Stars: \\Rigorous Formalism with Electrostatic Potential Perturbations}
\author{Andreas Reisenegger\altaffilmark{1}, Paula Jofr\'e\altaffilmark{1}, Rodrigo Fern\'andez\altaffilmark{2},
and Elena Kantor\altaffilmark{1,3}} \altaffiltext{1}{Departamento de
Astronom\'\i a y Astrof\'\i sica, Pontificia Universidad Cat\'olica
de Chile, Casilla 306, Santiago 22, Chile. E-mail:
areisene@astro.puc.cl.} \altaffiltext{2}{Department of Astronomy and
Astrophysics, University of Toronto, ON M5S 3H8, Toronto, Canada.}
\altaffiltext{3}{Ioffe Physical Technical Institute,
Politekhnicheskaya 26, 194021 Saint-Petersburg, Russia.}

\begin{abstract}
The electrostatic potential that keeps approximate charge neutrality
in neutron star matter is self-consistently introduced into the
formalism for \emph{rotochemical heating} presented in a previous
paper by Fern\'andez and Reisenegger. Although the new formalism is
more rigorous, we show that its observable consequences are
indistinguishable from those of the previous one, leaving the
conclusions of the previous paper unchanged.
\end{abstract}


\section{INTRODUCTION}
\label{sec:intoduction}

Rotochemical heating is one of several mechanisms that can keep
neutron stars (NSs) hot beyond their standard cooling time of $\sim
10^7~\mathrm{yr}$ (see \citealt{fernandez05} [hereafter FR05] for
further references, and \citealt{yakovlev04} for a recent review on
neutron star cooling). It originates in deviations from beta
equilibrium produced as the star gets compressed due to its
decreasing rotation rate, implying a conversion of rotational energy
into thermal energy. This effect is capable of keeping old
millisecond pulsars (MSPs) at surface temperatures $\sim 10^5$~K,
consistent with likely thermal ultraviolet emission from PSR
J0437-4720 (\citealt*{kargaltsev04}; FR05).

The heating mechanism works as follows. As a NS spins down, the
centrifugal force acting on each fluid element diminishes, raising
the local pressure. This compression results in a change of the
equilibrium concentration of each particle species. Reactions that
modify the chemical composition drive the system towards the new
equilibrium configuration. But if the rate at which reactions do
this task is slower than the change of the equilibrium
concentrations due to spin-down compression, the matter remains out
of chemical equilibrium. The excess energy stored in the chemical
imbalance is dissipated, partly by enhanced neutrino emission and
partly by heating the matter in the star. Thus, the star is kept
warm and can radiate for long after the standard cooling time.

After its introduction \citep{reisenegger95}, this heating mechanism
was studied by several authors: \citet{cheng96} applied it to quark
stars, \citet{reisenegger97} estimated the effects of superfluidity,
and \citet{iidasato97} studied the heating due to compositional
transitions in the crust due to spin-down compression.

The most striking prediction associated with rotochemical heating is
that, if the spin-down timescale is substantially longer than any
other timescale involved (with the likely exception of magnetic
field decay), the star arrives at a \emph{quasi-steady state}, in
which the temperature depends only on the current, slowly changing
value of $\Omega \dot{\Omega}$ (the product of the angular velocity
and its time derivative), and not on the star's previous history
\citep{reisenegger95}. This provides a simple way to constrain the
physics involved in theoretical models, once the spin parameters and
observed surface temperature of a given, sufficiently old NS are
known.

In spite of the previous work, FR05 were the first to go beyond
order-of-magnitude estimates and calculate the thermal evolution of
NSs with rotochemical heating, taking the structure of the star
fully into account in the frame of general relativity, and using
realistic equations of state of dense matter. They developed a
general formalism to treat the evolution of the temperature and
departures of chemical equilibrium, implementing it for the simplest
possible core composition, namely neutrons, protons, electrons, and
muons ($npe\mu$ matter). Using state-of-the-art equations of state,
and modified Urca reactions as the dissipative process, they
obtained results consistent with the observed ultraviolet emission
from the millisecond pulsar PSR J0437-4720 \citep{kargaltsev04}.

Here, we show that their formalism is not completely rigorous, in
the sense of ignoring the perturbations to the electrostatic
potential that must be present inside the NS in order to keep nearly
exact charge neutrality among the different charged particles
present inside the star.

We start (in \S 2) by discussing the physical origin, importance,
and strength of the electrostatic potential expected inside a NS, as
well as its relation to the (nearly exact) condition of local charge
neutrality. Then (in \S 3), we review the main components of the
rotochemical heating formalism developed by FR05, which are
rigorously correct and not directly affected by the presence of
electrostatic potential perturbations, showing that their main
result, namely the existence and properties of the quasi-steady
state, are unchanged. In \S 4, we introduce the electrostatic
potential into the equations of FR05 that relate the different
chemical imbalance variables (chemical potential imbalances and
deviations from equilibrium particle abundances) to each other. The
quantitative consequences of these modifications are explored in \S
5, whereas \S 6 contains our main conclusions.

\section{ELECTROSTATIC POTENTIAL IN A NEUTRON STAR}
\label{sec:electro}


It is well-known that a neutron ($n$) in vacuum is an unstable
particle, decaying into a proton ($p$), an electron ($e$), and an
electron antineutrino ($\bar\nu_e$). In NSs, neutrons are stabilized
by the presence of enough protons and electrons to block (through
the Pauli exclusion principle) the possible final states of the
decay. Neutron and protons have high masses and are therefore held
inside the star by its gravitational potential well, whose depth
must be larger than the kinetic part of their Fermi energies.
However, this potential is not deep enough to hold the much lighter
electrons, whose Fermi energy is relativistic (much larger than
their mass), and which would therefore tend to escape, yielding a
net positive charge on the NS. However, at NS densities, already a
small charge imbalance between protons and electrons causes an
electrostatic field that can hold in the electrons and at the same
time partially balance the gravitational force on the protons,
preventing them from sinking into the center of the star, due to
their relatively small kinetic energies. Thus, a non-rotating NS in
hydrostatic, chemical (beta), and diffusive equilibrium contains a
spherically symmetric electrostatic potential $\psi(r)$ that keeps
the charged particles near local charge neutrality. The existence of
such an electric potential in stars has been known to
astrophysicists since the early twentieth century. A recent
discussion, including early references, is found in
\citet{Neslusan}.

More formally, and using a relativistic description of gravity, in
contrast to the Newtonian description in the previous paragraph, we
describe the star by the spherically symmetric metric
\begin{equation} \label{eq:metric}
ds^2=-e^{2\Phi(r)}dt^2+e^{2\Lambda(r)}dr^2+r^2(d\theta^2+\sin^2\theta
d\phi^2)
\end{equation}
\citep{schutz}. The condition of diffusive equilibrium for each
particle species $i$ is that its total, redshifted chemical
potential,
\begin{equation} \label{eq:totalchem}
\mu_i^\infty=[\mu_i(r)+q_i\psi(r)]e^{\Phi(r)}
\end{equation}
is uniform throughout the star. Here, $q_i$ is the charge of each
particle species, and $\mu_i$ is its intrinsic chemical potential
\citep{kittel}, determined by the kinetic energy of the species (or
equivalently, by their number density $n_i$) and their local
interactions with particles of the same and other species (which are
determined by all the densities). For a given metric and
electrostatic potential, these conditions determine the distribution
of the individual particle species by fixing the spatial dependence
of the intrinsic chemical potential. For a typical density inside
NSs, $n_e\sim 10^{37}~\mathrm{cm^{-3}}$, the extremely relativistic
electron chemical potential $\mu_e\approx\hbar
c(3\pi^2n_e)^{1/3}\approx 10^2~\mathrm{MeV}$, which sets the scale
for the electrostatic potential, $|\psi|\sim\mu_e/e\sim
10^8~\mathrm{V}$, where $e\equiv q_p=-q_e$.

The field equation for the electrostatic potential is Maxwell's
equation, $F^{\alpha\beta}_{\phantom{\alpha\beta};\beta}=4\pi
J^\alpha$,
where the components of the electromagnetic field tensor
can be written in terms of the 4-vector potential components
$A^\beta$ as $F_{\alpha\beta}=A_{\beta;\alpha}-A_{\alpha;\beta}$,
and $J^\alpha$ are the components of the charge-current 4-vector. In
the static system of interest here (ignoring currents and magnetic
fields inside the neutron star), the space components of the
4-vectors can be set to zero: $A^i=0=J^i$, and the time component
$A^0\equiv\psi$. A substantial amount of algebra yields the
relativistic Poisson equation for the electrostatic potential,
\begin{equation} \label{eq:Poisson}
e^{-2\Lambda}\left[\psi''+\left(\Phi'-\Lambda'+{2\over
r}\right)\psi'\right]-e^{-2\Phi}R_{00}\psi=4\pi\sum_i n_iq_i,
\end{equation}
where $R_{\alpha\beta}$ are the components of the Riemann curvature
tensor, and the primes indicate derivatives with respect to the
radial coordinate $r$. The terms on the left-hand side (the
generalization of the laplacian operator of flat spacetime) are of
order $|\psi|/R^2$, where $R\sim 10^6~\mathrm{cm}$ is the radius of
the NS. Thus, a minute excess positive charge density
$\sim\mu_e/(eR)^2\sim 10^2~\mathrm{cm^{-3}}$ (compared to the
typical density of each charged species, $\sim
10^{37}~\mathrm{cm^{-3}}$, mentioned above) is enough to produce the
required potential. Therefore, to an excellent approximation, we can
assume that the electrostatic potential is exactly that required to
keep local neutrality everywhere among the charged particle species,
replacing equation~(\ref{eq:Poisson}) by the condition of charge
neutrality
\begin{equation} \label{eq:neutral}
\sum_i n_iq_i=0,
\end{equation}
imposed on the densities $n_i$, and obtaining the \emph{non-zero}
electrostatic potential from this condition, together with
equation~(\ref{eq:totalchem}).

\section{CHEMICAL IMBALANCE AND ROTOCHEMICAL HEATING}
\label{sec:heating}

The formalism of FR05 can be summarized as follows (we refer to the
original paper for more details). The changing rotation rate
$\Omega(t)$ of the star, assumed to be much slower than the
``break-up'' or ``Keplerian'' rate, slightly perturbs its structure,
making the total, chemical-equilibrium number $N_i^{eq}$ of
particles of species $i$ in the star change with time as
\begin{equation}
\label{eq:Nieq} \dot{N}_i^{eq} = 2 I_{\Omega, i}\Omega \dot{\Omega},
\end{equation}
where $I_{\Omega, i}$ is a constant, depending on the structure of
the unperturbed, non-rotating star. The actual number of particles
of the same species, $N_i$, is changed by non-equilibrium weak
interactions in the stellar core as
\begin{equation}
\label{eq:dot_Ni} \dot{N}_i = \int_{core} dV e^{\Phi}\sum_\alpha
\Delta \Gamma^i_\alpha.
\end{equation}
The integral is performed over the volume of the unperturbed,
non-rotating stellar core, and the sum runs over different kinds of
reactions (labeled by $\alpha$), each of which creates a net number
$\Delta \Gamma^i_\alpha$ of particles of species $i$ per unit time
and per unit volume. The latter quantity is a function of the local
density, temperature, and, most importantly, of the local chemical
imbalance, quantified by the variables
$\eta_{npl}=\mu_n-\mu_p-\mu_l$, where $l$ stands for the charged
leptons: electrons ($l=e$) or muons ($l=\mu$). If the $\eta_{npl}$
are non-zero, so are the net reaction rates, with their signs such
that the chemical imbalances tend to be decreased. The evolution of
the excess (or deficit) of particles of species $i$, $\delta
N_i\equiv N_i-N_i^{eq}$, is therefore determined by the competition
between the changing rotation rate of the star, which tends to
increase its absolute value, and the reactions, tending to decrease
it:
\begin{equation}
\label{eq:dot_dNi} {\delta\dot N}_i = \int_{core}
\left(e^{\Phi}\sum_\alpha \Delta \Gamma^i_\alpha\right) dV-2
I_{\Omega, i}\Omega \dot{\Omega}.
\end{equation}

The evolutionary time scale of the star is assumed to be much longer
than the time scales for diffusion of particles and heat across it.
So, in particular, the redshifted temperature
$T^\infty=T(r)e^{\Phi(r)}$ is considered to be uniform inside the
star, except for a thin surface layer. This temperature evolves as
\begin{equation} \label{eq:dot_T}
\dot{T}^\infty = \frac{1}{C}\left[ L^{\infty}_H - L^{\infty}_\nu -
L^{\infty}_\gamma \right],
\end{equation}
where $C$ is the total heat capacity of the star, $L^{\infty}_H$ is
the total power released by reactions and other heating mechanisms,
$L^{\infty}_\nu$ is the total neutrino luminosity, and
$L^{\infty}_\gamma$ is the photon luminosity, which are functions of
the chemical imbalances $\eta_{npl}$ and the temperature $T$
everywhere inside the star.

Although the particles are in diffusive equilibrium, moving easily
everywhere inside the star, they are generally not in chemical
equilibrium (free to convert into one another). Therefore, non-zero,
but spatially uniform, redshifted chemical imbalances are set up,
\begin{equation} \label{eq:imbalance}
\eta_{npl}^\infty\equiv\mu_n^\infty-\mu_p^\infty-\mu_l^\infty=\eta_{npl}e^\Phi,
\end{equation}
not involving the electrostatic potential, which cancels out due to
the equal and opposite charges of protons and leptons. Being uniform
in the star, these redshifted chemical imbalances are the ideal
variables to quantify the departure from chemical equilibrium and
follow its time evolution. The reaction rates on the right-hand side
of equations~(\ref{eq:dot_Ni}) and (\ref{eq:dot_dNi}) can be written
in terms of $\eta_{npl}^\infty$ and $T^\infty$ instead of the
corresponding un-redshifted, local, position-dependent variables.

In order to close the system of dynamical equations, it is necessary
to write the chemical imbalances, $\eta_{npl}^\infty$, in terms of
the excess numbers of particles, $\delta N_i$, which is where the
inaccuracy of the previous treatment (FR05) is located, as we shall
discuss below. However, at this point, we may note that, if we are
only interested in the quasi-steady state reached by NSs with slow
spin-down evolution, we only need to set the right-hand sides of
equations~(\ref{eq:dot_dNi}) and (\ref{eq:dot_T}) equal to zero,
immediately yielding a non-degenerate set of algebraic equations for
$\eta_{npl}^\infty$ and $T^\infty$, which is equivalent to that of
FR05 and therefore has identically the same solutions.

\section{RELATIONS AMONG THE VARIABLES QUANTIFYING THE CHEMICAL IMBALANCE}
\label{sec:imbalance}

When the chemical equilibrium is slightly perturbed, the total,
redshifted chemical potential of equation~(\ref{eq:totalchem}) is
displaced from its equilibrium value by the small amount
\begin{equation} \label{eq:dtotalchem}
\delta\mu_i^\infty=[\delta\mu_i+q_i\delta\psi+(\mu_i+q_i\psi)\delta\Phi]e^{\Phi},
\end{equation}
where the prefix $\delta$ refers to the respective variable being
slightly perturbed with respect to the chemical equilibrium state,
maintaining diffusive equilibrium (so $\delta\mu_i^\infty$ remains
uniform). We note that equation (9) of FR05 considered only the
first term. If some neutrons are converted into proton-electron
pairs (plus escaping antineutrinos), its effect on the energy
density and pressure, and therefore on the structure of the star and
the spacetime metric, is not large, as verified explicitly (in the
Newtonian approximation) in Appendix A. We therefore invoke the
relativistic Cowling approximation (e.~g., \citealt{Finn}), setting
$\delta\Phi=0$. However, the absolute value of the electrostatic
potential must increase, in order to hold more protons and electrons
together than before. This effect is considered explicitly in what
follows.

The intrinsic chemical potential perturbations are related to local
density changes, $\delta n_i=\sum_i(\partial
n_i/\partial\mu_j)\delta\mu_j,$ where, from
equation~(\ref{eq:dtotalchem}), $\delta\mu_j=\delta\mu_j^\infty
e^{-\Phi}-q_j\delta\psi$. Imposing conservation of local charge
neutrality, $\sum_iq_i\delta n_i=0$, yields
\begin{equation} \label{eq:dpsi}
\delta\psi={\sum_{k,m}q_k(\partial
n_k/\partial\mu_m)\delta\mu_m^\infty\over\sum_{r,s}q_rq_s(\partial
n_r/\partial\mu_s)}.
\end{equation}
Thus, after some algebra, the perturbation to the total number of
particles of one species in the star can be written as in equation
(11) of FR05:
\begin{equation} \label{eq:dNi}
\delta N_i=\int dV\delta n_i=\sum_jB_{ij}\delta\mu_j^\infty,
\end{equation}
but with the coefficients redefined as
\begin{equation} \label{eq:Bij}
B_{ij}=\int dV e^{-\Phi}\left[{\partial
n_i\over\partial\mu_j}-{\sum_{k,m}q_kq_m(\partial
n_i/\partial\mu_m)(\partial
n_k/\partial\mu_j)\over\sum_{r,s}q_rq_s(\partial
n_r/\partial\mu_s)}\right].
\end{equation}
(The second term inside the square brackets is missing in eq. [12]
of FR05.) The thermodynamic identity $\partial
n_i/\partial\mu_j=\partial n_j/\partial\mu_i$ can be used to show
that $B_{ij}=B_{ji}$. More important, however, is to note that the
$4\times 4$-matrix ${\bf B}$ is singular, since charge conservation
is ``built in,'' in the sense that, for any combination of four
values for the variables $\delta\mu_j^\infty$, it is guaranteed that
$\sum_iq_i\delta N_i=0$, i.~e., three of the rows of ${\bf B}$ are
not linearly independent: $B_{pj}=B_{ej}+B_{\mu j}$. One
manifestation of this singularity is the existence of ``trivial''
perturbations with $\delta\mu_i^\infty=q_i\delta\psi e^\Phi$, for
which $\delta\mu_i=0$ and therefore the particle distributions
inside the star are not changed. These perturbations can be
interpreted as redistributions of charges outside the star, or just
as redefinitions of the zero point of the potentials.

In order to obtain an invertible problem, we reduce our four degrees
of freedom to three, by defining the new variables
$\delta\tilde\mu^\infty_n\equiv\delta\mu^\infty_n$ and
$\delta\tilde\mu^\infty_l\equiv\delta\mu^\infty_p+\delta\mu^\infty_l,$
the latter for both $l=e,\mu$, and eliminating the row and column
corresponding to the protons from the matrix ${\bf B}$, in this way
converting it into a symmetric and invertible $3\times 3$-matrix
${\bf\tilde B}$, with $\delta N_i=\sum_{j=n,e,\mu}\tilde
B_{ij}\delta\tilde\mu^\infty_j$ for $i=n,e,\mu$. (The proton density
and total number can be recovered from those of electrons and muons
through the condition of charge neutrality.) The components of the
inverse matrix, ${\bf\tilde B}^{-1}$, relate the chemical imbalance
variables to the particle number perturbations:
$\eta^\infty_{npl}=\delta\tilde\mu^\infty_n-\delta\tilde\mu^\infty_l=\sum_{i=n,e,\mu}(\tilde
B_{ni}^{-1}-\tilde B_{li}^{-1})\delta N_i.$

Finally, this can be further reduced by considering that the
reactions also conserve baryon number\footnote{Note that, contrary
to charge neutrality, which can be imposed locally everywhere in the
star, baryon number is only a globally conserved quantity.}, i.~e.,
$\delta N_n=-\delta N_p=-\delta N_e-\delta N_\mu$, which allows to
also eliminate $\delta N_n$, writing
\begin{equation} \label{eq:eta2}
\eta^\infty_{npl}=\sum_{i=e,\mu}(\tilde B_{ni}^{-1}-\tilde
B_{li}^{-1}-\tilde B_{nn}^{-1}+\tilde B_{ln}^{-1})\delta N_i.
\end{equation}
We can make contact with the notation of FR05 by writing
\begin{equation} \label{eq:etaZ}
\eta^\infty_{npe}=-Z_{npe}\delta N_e-Z_{np}\delta
N_\mu\qquad\mathrm{and}\qquad\eta^\infty_{np\mu}=-Z_{np}\delta
N_e-Z_{np\mu}\delta N_\mu,
\end{equation}
with the coefficients rewritten as
\begin{equation} \label{eq:Znp}
Z_{np}\equiv\tilde B_{nn}^{-1}-\tilde B_{ne}^{-1}-\tilde
B_{n\mu}^{-1}+\tilde B_{e\mu}^{-1},
\end{equation}
\begin{equation} \label{eq:Znpe}
Z_{npe}\equiv\tilde B_{nn}^{-1}-2\tilde B_{ne}^{-1}+\tilde
B_{ee}^{-1},
\end{equation}
\begin{equation} \label{eq:Znpm}
Z_{np\mu}\equiv\tilde B_{nn}^{-1}-2\tilde B_{n\mu}^{-1}+\tilde
B_{\mu\mu}^{-1},
\end{equation}
instead of equations (54) through (56) of FR05. These re-definitions
were chosen so that their equations (52) and (53) for the time
evolution of the chemical imbalance variables $\eta^\infty_{npl}$
remain formally correct. The coefficients $W_{npl}$, which determine
the rate by which chemical imbalances are built up by the changing
rotation rate, also remain correctly expressed by equations (57) and
(58), although more easily written as
\begin{equation} \label{eq:Wnpl}
W_{npe}=Z_{npe}I_{\Omega,e}+Z_{np}I_{\Omega,\mu}\qquad\mathrm{and}\qquad
W_{np\mu}=Z_{np}I_{\Omega,e}+Z_{np\mu}I_{\Omega,\mu}.
\end{equation}

\section{QUANTITATIVE EVALUATION AND DISCUSSION}
\label{sec:discussion}

Formally, the only changes that have to be introduced to the
equations of FR05 are to replace their equations (9), (12), (13),
(54), (55), and (56) by the corresponding equations in the previous
sections. In their dynamical system for the time evolution of the
variables $T^\infty$ and $\eta^\infty_{npl}$, the only difference
lies in the values of the coefficients $Z_{np}$, $Z_{npe}$, and
$Z_{np\mu}$, now defined by our equations (\ref{eq:Znp}),
(\ref{eq:Znpe}), and (\ref{eq:Znpm}), instead of their equations
(54), (55), and (56), and in the coefficients $W_{npl}$, which
depend on the former (e.~g., our eq. [\ref{eq:Wnpl}]). We have
argued in \S 3 that the quasi-steady state eventually reached by old
NSs is unaffected by these changes. However, the rate at which this
state is approached (directly depending on the coefficients
$W_{npl}$; see, e.~g., FR05, eqs. [76] and [81]) could be affected,
as well as the less intuitive time evolution of younger NSs with
stronger magnetic fields (e.~g., \citealt{reisenegger95}).
\clearpage
\begin{figure}
\plotone{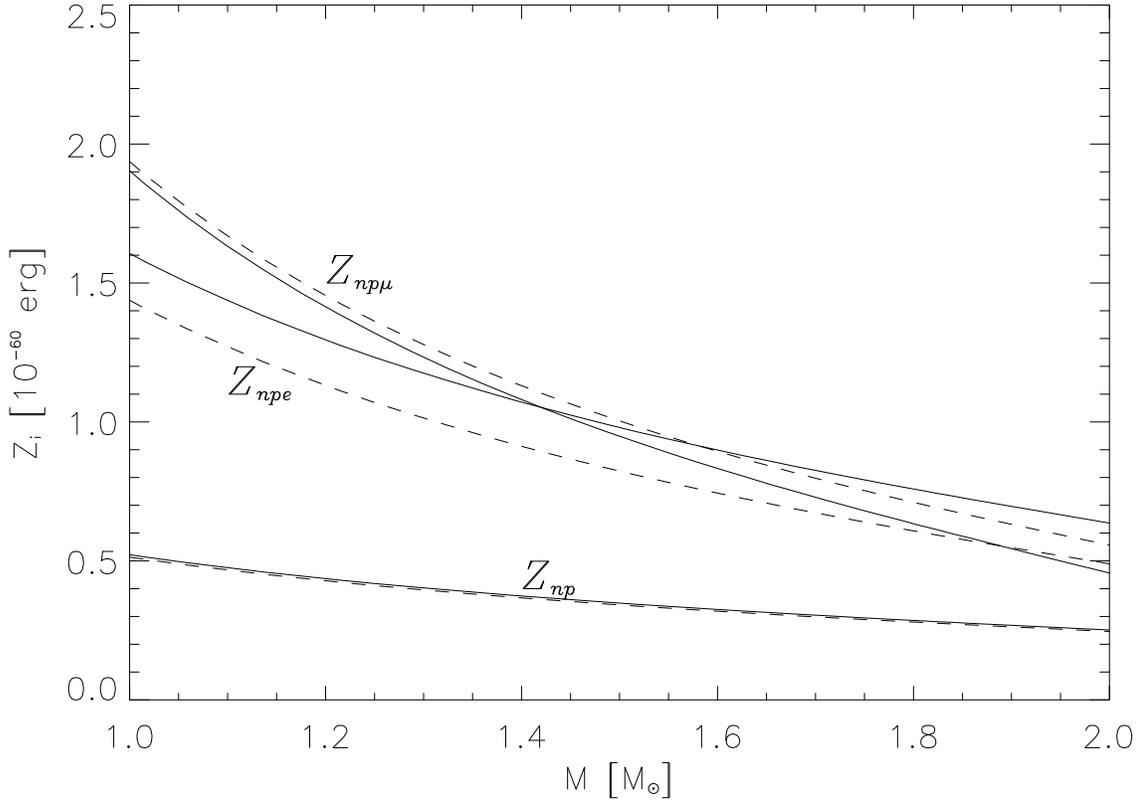} \caption{Coefficients $Z_{np}$, $Z_{npe}$, and
$Z_{np\mu}$ as functions of stellar mass, for models built with the
A18 + $\delta \upsilon$ + UIX* equation of state \citep{apr98}.
Dashed lines correspond to the formalism of FR05 (without explicit
account of electrostatic potentials), while solid lines are the
results of the corrected formalism presented in the present paper.
\label{fig:Z}}
\end{figure}
\clearpage
Figure~\ref{fig:Z} shows a comparison of the coefficients $Z_{np}$,
$Z_{npe}$, and $Z_{np\mu}$, as functions of stellar mass, for one
particular equation of state (A18+$\delta \upsilon$+UIX*,
\citealt{apr98}), for the formalism of FR05 (dashed lines) and the
more rigorous one presented here (solid lines). It can be seen that
$Z_{np}$ hardly changes, whereas $Z_{npe}$ and $Z_{np\mu}$ show
changes whose fractional amplitude increases with mass, up to a
$\sim 10-20~\%$ level at the largest masses. Calculations with other
equations of state yielded very similar results.
\clearpage
\begin{figure}
\plottwo{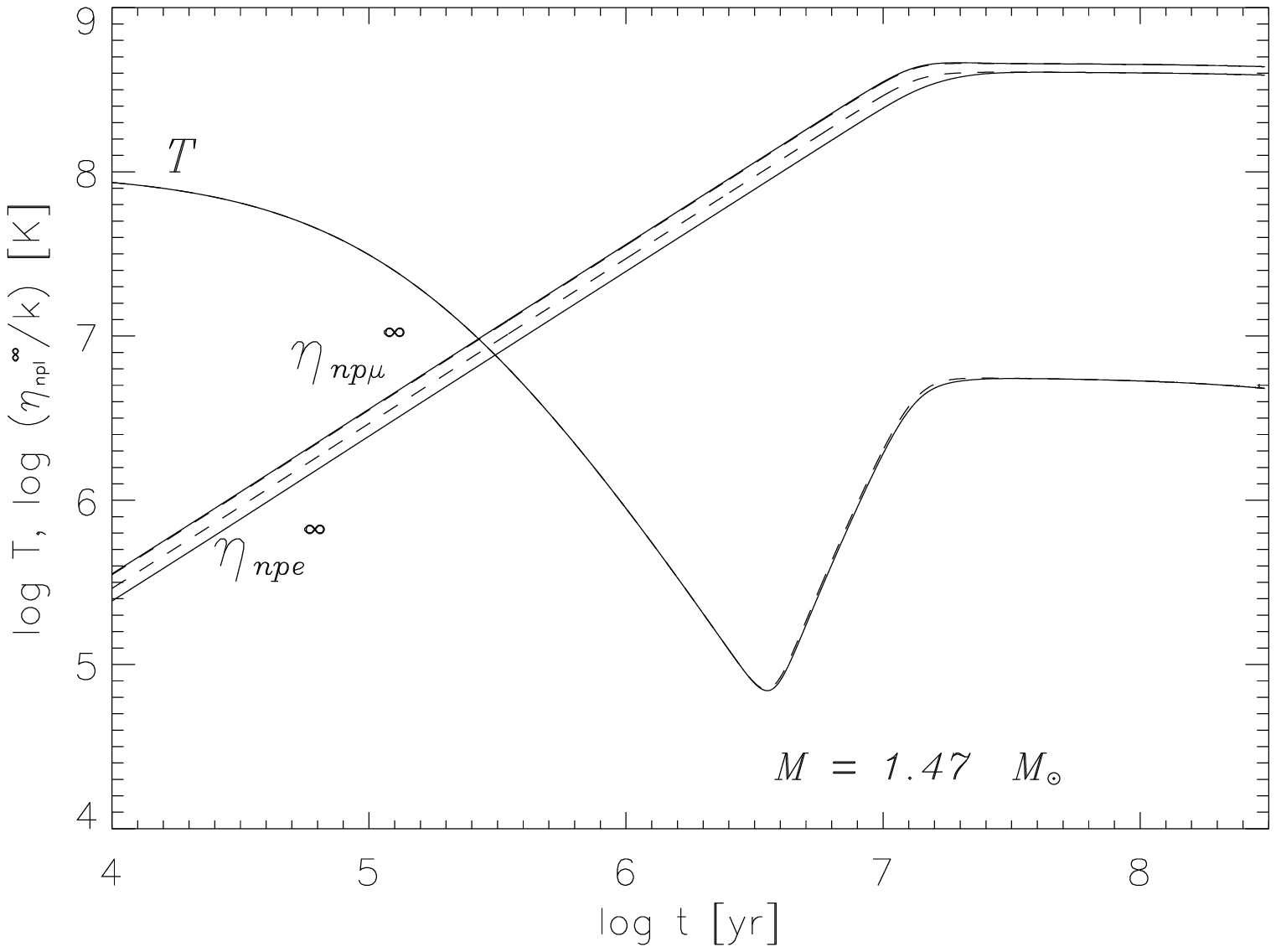}{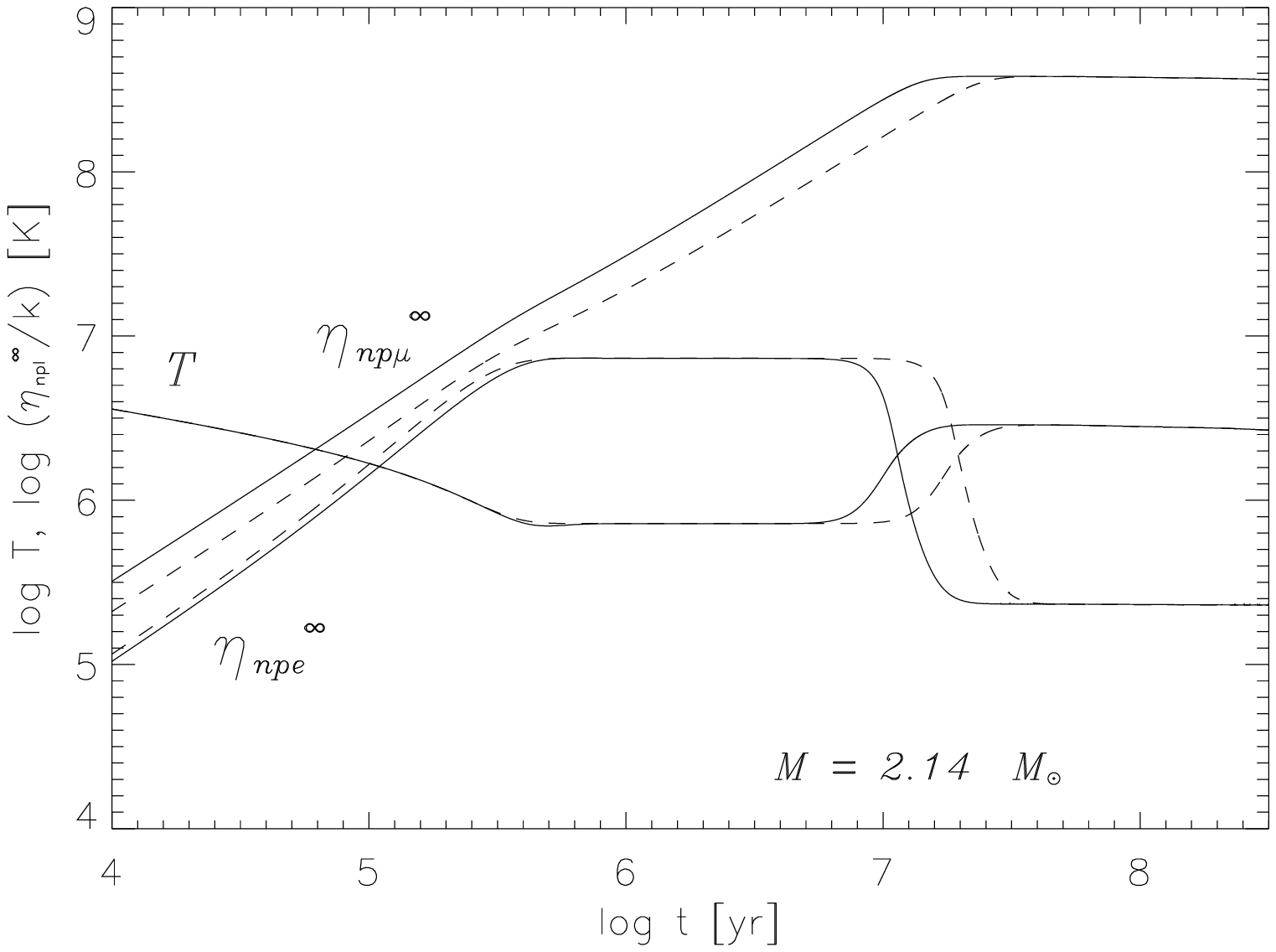} \caption{Evolution of the internal
redshifted temperature, $T^\infty$, and chemical imbalance variables
$\eta_{npe}^\infty/k$ and $\eta_{np\mu}^\infty/k$, for NS models of
$1.47~M_{\odot}$ (left) and $2.13~M_{\odot}$ (right), calculated
with the A18 + $\delta \upsilon$ + UIX* equation of state
\citep{apr98}, with initial conditions of vanishing chemical
imbalances and $T_{\infty} = 10^8$ K at $t=0$, and assuming magnetic
dipole spin-down with field strength $B = 10^8$ G and initial period
$P_0 = 1$ ms. Dashed lines represent results for the formalism of
FR05, whereas solid lines correspond to the present, corrected
formalism.
 \label{fig:evolution}}
\end{figure}
\clearpage
In Figure~\ref{fig:evolution}, we compare the time evolution of the
chemical imbalance variables, $\eta_{npe}^\infty$ and
$\eta_{np\mu}^\infty$, as well as the redshifted internal
temperature, $T^\infty$, as obtained in both formalisms for two
different neutron star masses. For the lower mass, particles inside
the star undergo only (slow) modified Urca reactions, whereas at the
higher mass electrons can undergo (fast) direct Urca processes,
whereas muons are still restricted to modified Urca. This accounts
for the relatively simple evolutionary curves on the left, as
compared to the more complex one on the right. (See FR05 for a more
detailed discussion.) It can be seen that, consistent with our
previous discussion, there is no evidence for differences between
the two formalisms in the asymptotic, quasi-steady regime, whereas
some differences are seen at the epochs of strong time evolution,
especially in the higher mass star.

\section{CONCLUSIONS}
\label{sec:conclusions}

We have shown that the electrostatic potential inside a NS plays an
important role in determining the spatial distribution of charged
particles and their nearly exact local charge neutrality everywhere
in the star. We take it into account explicitly in an extension of
the formalism for rotochemical heating of FR05, in which the
relations between the two sets of variables quantifying chemical
imbalances (chemical potential differences and departures from
equilibrium abundances) are affected, without changing the rest of
the formalism. The quasi-steady state reached by old NSs is strictly
unaffected by these corrections, and the time-dependent evolution at
earlier times shows quantitatively negligible changes.

\acknowledgments The authors are very grateful to Michael Gusakov
for pointing out the problem that we address in the present paper.
We also thank Olivier Espinosa for useful discussions, and Juan
V\'eliz for assistance with the preparation of figures. This work
received financial support from FONDECYT Regular Grants 1020840 and
1060644, from the FONDAP Center for Astrophysics, from the Visiting
Professors Program of Pontificia Universidad Cat\'olica de Chile,
from the Russian Leading Scientific School grant 9879.2006.2, and
from the Russian Fund for Basic Research grant 04-02-17590.

\appendix

\section{Quantitative evaluation of gravitational potential perturbations}
\label{sec:gravpot}
\clearpage
\begin{figure}
\plotone{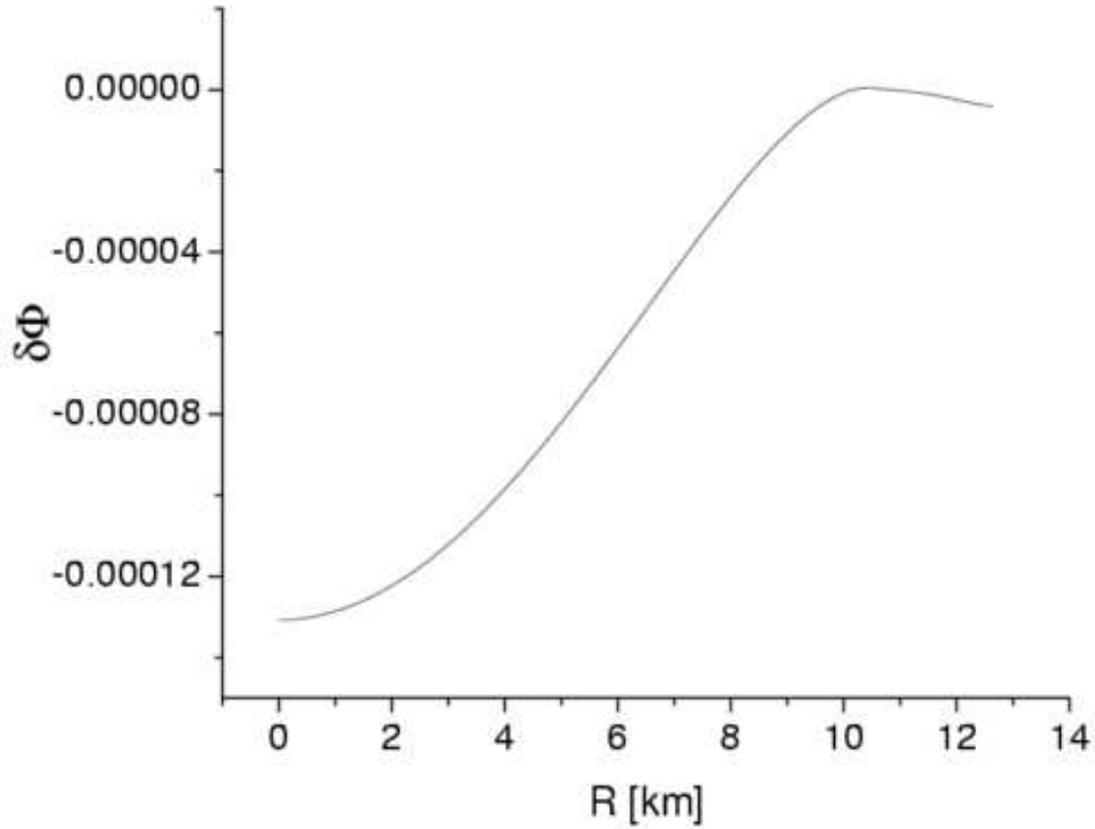} \caption{Perturbation of the gravitational
potential in a neutron star model with the first PAL equation of
state \citep{pal88} and total mass $M=1.6~M_\odot$, subject to a
chemical perturbation corresponding to a 1\% increase of the
electron and proton density everywhere in the star, together with a
corresponding decrease in the neutron density, so as to conserve
total baryon number. \label{fig:gravpot}}
\end{figure}
\clearpage
Here we verify whether the gravitational potential (or metric)
perturbation in equation (\ref{eq:totalchem}) can indeed be
neglected. We consider a neutron star model of $1.6~M_\odot$,
calculated in full General Relativity with the first PAL equation of
state \citep{pal88}. For simplicity using perturbation theory in a
Newtonian framework, we calculate the gravitational potential
perturbation $\delta\Phi$ corresponding to an arbitrarily chosen
deviation from chemical equilibrium, taken as an excess electron and
proton density of 1\% everywhere (muons are not present in this
model), with a corresponding decrease in the neutron density, so as
to conserve baryon number. Figure~\ref{fig:gravpot} shows
$\delta\Phi$ as a function of the radial coordinate. For comparison,
$\delta\mu_e/\mu_e\approx\delta n_e/(3n_e)\approx 0.003$, so the
intrinsic chemical potential perturbation term in equation
(\ref{eq:totalchem}) is about 50 times larger than that
corresponding to the gravitational potential.




\end{document}